  \documentstyle[12pt]{article}
  
\catcode`@=11
\def\secteqno{\@addtoreset{equation}{section}%
\def\theequation{\thesection.\arabic{equation}}}
\catcode`@=12
\secteqno
\newcommand{\be}{\begin{equation}}
\newcommand{\ee}{\end{equation}}

\def\a{\alpha}              \def\g{\gamma}   \def\d{\delta}
\def\e{\epsilon}         
             
                         \def\p{\pi}
       \def\s{\sigma}             
   \def\v{\psi}             

\newcommand{\fa}{\frac{1}{\alpha}}
\newcommand{\fat}{\frac{2}{\alpha}}

\def\V{\Psi}               
\newcommand{\nn}{\nonumber}

\begin{document}
          \hfill 

	  \hfill March, 2002

	  \hfill hep-lat/0203019

          \hfill NIIG-DP-02-2 
\vskip 20mm

\begin{center} 
{\bf \Large Ginsparg-Wilson Relation and Lattice Chiral
Symmetry\\ in Fermionic Interacting Theories}

\vskip 10mm
{\large Yuji\ Igarashi$^a$, Hiroto\ So and Naoya Ukita}\par

\medskip
{\it 
$^a$ Faculty of Education, Niigata University, Niigata 950-2181, Japan\\
Department of Physics, Niigata University, Niigata 950-2181, Japan\\
}

\medskip
\date{\today}
\end{center}
\vskip 10mm
\begin{abstract}
We derive Ginsparg-Wilson relation for a lattice chiral symmetry in
 theories with self-interacting fermions.  Auxiliary scalar and
 pseudo-scalar fields are introduced on a coarse lattice to give an
 effective description of the fermionic interactions. We obtain
 particular solutions to the Ginsparg-Wilson relation and other
 Ward-Takahashi identities in a closed form. These non-perturbative
 solutions can be used to construct a chiral invariant action and
 an invariant path-integral measure on the coarse lattice.  The resulting
 partition function exhibits the exact chiral symmetry in the fermionic
 system with the auxiliary fields.

\end{abstract}

\newpage
\setcounter{page}{1}
\setcounter{footnote}{0}
\parskip=7pt

\section{Introduction}

In the last decade, considerable progress has been made in realization
of the chiral symmetry on the lattice. The crucial issue in this realization
lies in an algebraic constraint on lattice Dirac operators derived by
Ginsparg and Wilson \cite{gw}.  An explicit solution to the
Ginsparg-Wilson (GW) relation \cite{gw} was given \cite{neu1}. It
exhibits some remarkable properties concerning chirality
\cite{hln}\cite{lu1}\cite{has} and locality
\cite{her}\cite{neu2}\cite{Hor}: the Dirac operator is not ultra-local
but exponentially local.  In the L\"{u}scher's realization of the chiral
symmetry \cite{lu1}, the Dirac operator appears in the symmetry
transformations, thereby avoiding the no-go theorem \cite{nn}.  It has
been also shown that this formulation of the chiral symmetry correctly gives
the index theorem related to the chiral anomaly \cite{hln}\cite{lu1}.
See, for example, ref.\cite{reviews} for reviews of recent development.

The above considerations of the chiral symmetry are based on the GW
relation which is derived mainly in free theories.  It is highly
nontrivial how lattice chiral symmetry can be realized in theories
with generic interactions.  In order to formulate the symmetry, one has
to construct symmetry transformations in such a way that both of action
and path-integral measure become chiral invariant. Recently, L\"{u}scher
has shown \cite{lu2} that this can be done in abelian chiral gauge
theories with anomaly-free fermion multiplets. This remarkable success
is tempting us to formulate lattice chiral symmetry in other interacting
theories.

The purpose of this letter is to discuss chiral symmetry in fermionic
self-interacting system.  To formulate the symmetry, we begin with a
microscopic fermionic action on a fine lattice, and then define a
macroscopic action on a coarse lattice using the block-spin
transformation.  The path-integral over the microscopic fields will
generate fermionic self-interaction terms in the macroscopic
action. Instead of dealing with such terms directly, we introduce
auxiliary fields on the coarse lattice to give an effective description
of the fermionic interaction terms.  We restrict ourselves here to
scalar and pseudo-scalar fields. The lattice action consists of
potential of these fields, in addition to fermion bilinear forms with
the Yukawa coupling.  Our main task is then to derive the Ward-Takahashi
(WT) identities which includes the GW relation for this action. The
resulting GW relation depends on the auxiliary fields, and gives an
extension of the original GW relation in free theories.  The relation we obtain
can be solved by making suitable locality assumption on the chiral
transformations of the auxiliary fields. It is found in our construction
that the auxiliary fields are transformed nonlinearly. We determine
chiral invariants made of the auxiliary fields, and a counter term
needed to cancel the change in the path integral measure defined on the
coarse lattice.  This can be done by solving the remaining WT
identities. The non-perturbative solutions we obtain can be used to
define an exact chiral symmetry in quantum system with the auxiliary
fields describing the scalar and pseudo-scalar couplings of the
fermionic self-interactions. Our approach should be compared with those
in earlier attempts \cite{lu1}\cite{cwz} \cite{in}\cite{iis}, which were
based on the GW relation in free theories.

Since our construction of the symmetry
is slightly involved, we present here our main results. A detailed
description of the chiral symmetry and its structure will be given
elsewhere \cite{isu}.

\section{Macroscopic action with auxiliary fields}
Let us consider a microscopic action $A_{\rm c}[\v_{\rm c},\bar{\v}_{\rm
c}]$ of the Dirac fields $\v_{\rm c}(x),\bar{\v}_{\rm c}(x)$ carrying a
single flavor.  These fields are defined on a $d$ (even) dimensional
fine-lattice whose positions are labeled by $x$.  The microscopic
action describes a certain class of fermionic self-interactions
specified below. The action is assumed to be invariant under the chiral
transformation
\begin{eqnarray}
 \left\{
 \begin{array}{ccc}
   \v_{\rm c}(x) & \longrightarrow & 
    \v_{\rm c}^{\prime}(x) \equiv \v_{\rm c}(x) + i\e\g_{5}\v_{\rm c}(x) \\
   \bar{\v}_{\rm c}(x) & \longrightarrow & 
    \bar{\v}_{\rm c}^{\prime}(x) \equiv \bar{\v}_{\rm c}(x)                                   + \bar{\v}_{\rm c}(x)i\e\g_{5} \\
 \end{array}
 \right. ,
 \label{2.1}
\end{eqnarray}
with an infinitesimal constant parameter $\e$: 
\begin{eqnarray}
 A_{\rm c}[\v_{\rm c},\bar{\v}_{\rm c}] & \longrightarrow &
 A_{\rm c}[\v_{\rm c}^{\prime},\bar{\v}_{\rm c}^{\prime}]
 = A_{\rm c}[\v_{\rm c},\bar{\v}_{\rm c}] .
\label{2.2}
\end{eqnarray}
Let $A[\V,\bar{\V}]$ be an effective action of the Dirac fields
$\V_{n},\bar{\V}_{n}$ defined on a coarse lattice.
Indices $n, m$ are used for labeling sites of the lattice.  
This macroscopic action is obtained from the microscopic action 
via the block-spin transformation \cite{gw} 
\begin{eqnarray}
 e^{-A[\V,\bar{\V}]} & \equiv &
 \int{\cal D}\v_{\rm c}{\cal D}\bar{\v}_{\rm c}\ 
 e^{- A_{\rm c}[\v_{\rm c},\bar{\v}_{\rm c}]
    - \sum_{n}(\bar{\V}_{n}-\bar{B}_{n})\a(\V_{n}-B_{n})},
 \label{2.3}
\end{eqnarray}
where $\a$ is a constant parameter proportional to inverse of the coarse
lattice spacing $a$, $\a \propto 1/a$. The gaussian integral in (\ref{2.3})
relates the macroscopic fields $\V_{n},\bar{\V}_{n}$ to the block-spin
variables defined by
\begin{eqnarray}
 \left\{
 \begin{array}{ccl}
   B_n & \equiv &
   \int d^{d}x\ f_n(x) \v_{\rm c}(x) \\
   \bar{B}_n & \equiv &
   \int d^{d}x\ \bar{\v}_{\rm c}(x) f_{n}^{\ast}(x)
 \end{array}
 \right. ,
 \label{2.4}
\end{eqnarray}
where $f_{n}(x)$ is an appropriate 
function for coarse graining normalized as 
$\int d^{d}x\ f_{n}^{\ast}(x) f_{m}(x) = \d_{nm}$.

The path-integral over the microscopic fields in (\ref{2.3}) will
generate fermionic self-interaction terms in the macroscopic action
$A[\V,\bar{\V}]$. Dealing with such terms directly makes our symmetry
analysis complicated.  We consider instead an effective theory with some
auxiliary fields, where the fermionic interaction terms generated by the
block-spin transformation can be replaced by an appropriate function of
the auxiliary fields. We introduce auxiliary fields $\s_{n}$ and
$\p_{n}$ on the coarse lattice to describe the fermionic scalar and
pseudo-scalar interactions. As is well known, these interactions are
incorporated into the Nambu-Jona-Lasino model, and recognized as the
most important couplings to describe the chiral symmetry and its spontaneous
breaking in the effective theory. Therefore, it is reasonable to
consider these auxiliary fields, and inclusion of other fields can be
done in a similar manner.

The macroscopic action we consider then takes of the form
\begin{eqnarray}
 A[\V,\bar{\V}] = \Sigma_{nm}\bar{\V}_{n} (D_{0})_{nm} \V_{m} + 
 V[\bar{\V}_{n} (\d_{nm} + h(\nabla)_{nm})\V_{m}, \bar{\V}_{n} \gamma_{5}
 (\d_{nm} + h(\nabla)_{nm})\V_{m}] ,
\label{2.5}
\end{eqnarray}
where $D_{0}$ is the Dirac operator for the kinetic term, and $V$ denotes 
fermionic interactions which consist of contact terms as well as 
non-contact ones with the difference operators $h(\nabla)_{nm}$.
We may obtain the action (\ref{2.5}) by performing integration 
over the auxiliary fields in a new macroscopic action:
\begin{eqnarray}
 e^{-A[\V,\bar{\V}]} & \equiv &
 \int{\cal D}\p{\cal D}\s   
 e^{- \sum_{nm}\bar{\V}_{n}
 \left(\left(D_{0}\right)_{nm} +\, \left(\d + h(\nabla)\right)_{nm}
 (i\g_5\p + \s )_{m}\right)\V_{m}- A_{X}[\p,\s]}.
 \label{2.6}
\end{eqnarray}
It is noted that the Dirac fields appear only bilinearly in the new action. 
All the fermionic interactions are cast into the Yukawa couplings
 with the auxiliary fields and the potential term $A_{X}[\p,\s]$.  
The new action given above can be used to construct an exact chiral symmetry 
in the macroscopic theory.

\section{Ginsparg-Wilson relation in fermionic interacting theory}
We derive now GW relation for a chiral symmetry in the fermionic
interacting theory on the coarse lattice. Our derivation is based on the
path-integral relation obtained above:
\begin{eqnarray}
 \lefteqn{\int{\cal D}\p{\cal D}\s \  
  e^{- \sum_{nm}\bar{\V}_{n}{\tilde D}(\p,\s)_{nm} \V_{m}
     - A_{X}[\p,\s]}} & & \nonumber \\ 
 & \hspace{2cm} = & 
 \int{\cal D}\v_{\rm c}{\cal D}\bar{\v}_{\rm c}\  
  e^{- A_{\rm c}[\v_{\rm c},\bar{\v}_{\rm c}]
     - \sum_{n}(\bar{\V}_{n}-\bar{B}_{n})\a(\V_{n}-B_{n})} ,
 \label{2.7}
\end{eqnarray}
where the total Dirac operator is given by
\begin{eqnarray}
{\tilde D}(\p,\s)_{nm} &=& (D_{0})_{nm} + \left(\d + h(\nabla)\right)_{nm}
(i\g_5\p + \s)_{m} \nonumber\\
&\equiv& D(\p,\s)_{nm} + \d_{nm}(i\g_5\p + \s)_{m}.    
\label{2.8}
\end{eqnarray}
It is assumed that the ${\tilde D}(\p,\s)$ is at most linear in $\pi$
and $\s$. 

In oder to obtain the GW relation as a WT identity, 
we make a change of variables in both sides of the path-integral. For
the microscopic fields $\v_{c}(x),\bar{\v}_{c}(x)$, we use the
chiral transformation given in (\ref{2.1}) to have new variables.  
They lead to new block-spin variables,
\begin{eqnarray}
 \left\{
 \begin{array}{c}
    B_{n}^{\prime} \equiv B_n + i\e\g_{5}B_n \\
    \bar{B}_{n}^{\prime} \equiv \bar{B}_{n} 
                                  + \bar{B}_{n}i\e\g_{5} 
 \end{array}
 \right. .
 \label{2.9}
\end{eqnarray}
For the macroscopic fields $\V_n,\bar{\V}_n$, it is reasonable to consider 
\begin{eqnarray}
 \left\{
 \begin{array}{ccc}
    \V_{n}^{\prime} \equiv \V_{n} + i\e\g_{5}\V_{n} \\
    \bar{\V}_{n}^{\prime} \equiv \bar{\V}_{n} 
                                  + \bar{\V}_{n}i\e\g_{5}
 \end{array}
 \right. .
 \label{2.10}
\end{eqnarray}
We may write the transformation on the auxiliary fields as
\begin{eqnarray}
 \left\{
 \begin{array}{ccc}
    \p_{n}^{\prime} \equiv \p_{n} + \e \ \d\p_{n} \\
    \s_{n}^{\prime} \equiv \s_{n} + \e \ \d\s_{n}
   \label{2.11}  
 \end{array}
 \right. .
\end{eqnarray}
where $\d\p_n$ and $\d\s_n$ are unknown, and to be determined below.
For the Dirac operator $\tilde{D}(\p,\s)$, we have
\begin{eqnarray}
 \tilde{D}^{\prime} \equiv
 \tilde{D} + \e\, \d D + \e \, (i\g_5\d\p + \ \d\s) .
 \label{2.12}
\end{eqnarray}

Invariance of the path-integral under the above change of variables
gives the WT identity for our interacting system. It is derives as
follows. Let us make the change of variables given in (\ref{2.1}),
(\ref{2.9}), (\ref{2.10}) and (\ref{2.11}).  Then, the l.h.s of
(\ref{2.7}) expressed by new variables turns to be
\begin{eqnarray}
 \lefteqn{ 
  \int{\cal D}\p^{\prime}{\cal D}\s^{\prime} \  
   e^{- \sum_{nm}\bar{\V}^{\prime}_{n}
                 \tilde{D}(\p^{\prime},\s^{\prime})_{nm}
                 \V^{\prime}_{m}
      - A_{X}[\p^{\prime},\s^{\prime}]}}  \nonumber \\ 
 \nonumber \\
 & = &
   \int{\cal D}\p{\cal D}\s  \nonumber \\
 &   &
   \times\left[
    1 + \e\left(\d J_{X}-\d A_{X}\right)
      - \e\sum_{nm}\bar{\V}_n\left(i\{\g_5,\tilde{D}\}
                                   +\d \tilde{D}\right)_{nm}\V_m
      + {\cal O}(\e^2) 
   \right]  \nonumber \\
 &   &
   \times \exp\left(- \sum_{nm}\bar{\V}_{n}\tilde{D}_{nm}\V_{m}
      - A_{X}\right)  ,
 \label{a2}
\end{eqnarray}
where $\d J_{X}$ denotes the change of 
${\cal D}\p{\cal D}\s$.  $\d A_{X}$ as well as $\d \tilde{D}$ are 
induced by $\d \pi$ and $\d \s$.  On the other hand, the r.h.s of (\ref{2.7}) turns to be 
\begin{eqnarray}
 \lefteqn{ 
   \int{\cal D}\v^{\prime}_{\rm c}{\cal D}\bar{\v}^{\prime}_{\rm c}\  
    e^{- A_{\rm c}[\v^{\prime}_{\rm c},\bar{\v}^{\prime}_{\rm c}]
      - \sum_{n}(\bar{\V}^{\prime}_{n}-\bar{B}^{\prime}_{n})\a
                (\V^{\prime}_{n}-B^{\prime}_{n})}}  \nonumber \\
 \nonumber \\
 & = &
  \int{\cal D}\v_{\rm c}{\cal D}\bar{\v}_{\rm c} \nonumber \\
 &   &
  \times\left[
   1 + \e \left(\d J_{\rm c}- \d A_{c} 
     - \sum_{n}(\bar{\V}_{n}-\bar{B}_{n})2i\a\g_5(\V_{n}-B_{n})\right)  
     + {\cal O}(\e^2)
  \right] \nonumber \\  
 &   &
  \times \exp\left(- A_{\rm c}[\v_{\rm c},\bar{\v}_{\rm c}]
      - \sum_{n}(\bar{\V}_{n}-\bar{B}_{n})\a(\V_{n}-B_{n})\right) 
  \nonumber \\ 
 \nonumber \\
 & = &
  \int{\cal D}\v_{\rm c}{\cal D}\bar{\v}_{\rm c} \nonumber \\
 &   &
  \times\left[
   1 + \e\,(\d J_{\rm c}- \d A_{c}) 
     +\e \sum_{n}\left(\frac{2i}{\a}\frac{\partial^{l}}{\partial \V_n}
                        \g_5\frac{\partial^{l}}{\partial \bar{\V}_n}
                        +2i{\rm Tr}\g_5\right)  
     + {\cal O}(\e^2)
  \right] \nonumber \\  
 &   &
  \times \exp\left(- A_{\rm c}[\v_{\rm c},\bar{\v}_{\rm c}]
      - \sum_{n}(\bar{\V}_{n}-\bar{B}_{n})\a(\V_{n}-B_{n})\right),  
\label{a3}
\end{eqnarray}
where $\d J_{\rm c}$ originates from the change of ${\cal D}\v_{\rm
c}{\cal D}\bar{\v}_{\rm c}$. The third term with square brackets in the
last expression of (\ref{a3}) can be placed outside of the integral over the
microscopic variables. It leads to
\begin{eqnarray} 
\lefteqn{
\e \sum_{n}\left(\frac{2i}{\a}\frac{\partial^{l}}{\partial \V_n}
                        \g_5\frac{\partial^{l}}{\partial \bar{\V}_n}
                        +2i{\rm Tr}\g_5\right) 
                        \int{\cal D}\p{\cal D}\s \  
  e^{- \sum_{nm}\bar{\V}_{n}\tilde{D}(\p,\s)_{nm}\V_{m}
     - A_{X}[\p,\s]}} \nonumber \\ \nn \\
 & = & \int{\cal D}\p{\cal D}\s \                          
  \e \left[2i\sum_{n}{\rm Tr}\left(\g_5
              -\g_5\frac{\tilde{D}}{\a}\right)_{nn}- 
              \sum_{nm}\frac{2i}{\a}\bar{\V}_n
                           (\tilde{D}\g_5\tilde{D})_{nm}\V_m  \right]
                   \nonumber \\ 
 &   &
  \times
  \exp\left(- \sum_{nm}\bar{\V}_{n}\tilde{D}(\p,\s)_{nm}\V_{m}
     - A_{X}[\p,\s]\right) .
\label{a4}
\end{eqnarray}  
Collecting the contributions of order 
${\cal O}(\e)$ in (\ref{a2}) and (\ref{a3}), we obtain     
\begin{eqnarray} 
\lefteqn{
 \int{\cal D}\v_{\rm c}{\cal D}\bar{\v}_{\rm c}\  
(\d J_{\rm c}- \d A_{c})\ e^{- A_{\rm c}[\v_{\rm c},\bar{\v}_{\rm c}]
      - \sum_{n}(\bar{\V}_{n}-\bar{B}_{n})\a(\V_{n}-B_{n})}}
  \nonumber \\ 
&=& \int{\cal D}\p{\cal D}\s  \biggl[ \left(\d J_{X}-\d A_{X}
\right) - 
  2i\sum_{n}{\rm Tr}\left(\g_5
              -\g_5\frac{\tilde{D}}{\a}\right)_{nn}\nonumber\\
&{}& \qquad \qquad \ \ 
 - \sum_{nm}\bar{\V}_n\left(i\{\g_5,\tilde{D}\}
                                   +\d \tilde{D}- \frac{2i}{\a}
                           \tilde{D}\g_5\tilde{D}\right)_{nm}\V_m
 \biggr] \nonumber \\  
 &{}& \times   \exp\left(- \sum_{nm}\bar{\V}_{n}\tilde{D}(\p,\s)_{nm}\V_{m}
     - A_{X}[\p,\s]\right).
\label{2.13}
\end{eqnarray}
It should be noted that since one introduces the auxiliary fields only
on the coarse lattice, one has to perform their integration in the
macroscopic action.  
 
For the l.h.s. of (\ref{2.13}), we may take $\d A_{c} = \d J_{\rm c}
=0$.  The former follows from our requirement of (\ref{2.2}).  The
latter, the condition of absence of anomaly in the microscopic theory,
is expected to be the case because the theory has no gauge fields. Turn
to the r.h.s. of (\ref{2.13}), therefore, the path-integral over the
auxiliary fields gives zero. This allows a wide class of solutions for
which the integrand becomes $\pi$ or $\s$ derivative of something.  We
consider here more restrict class of solutions for which the integrand
itself vanishes. In order to further reduce this condition, we assume
that $\d \p_{n}$ and $\d \s_{n}$ are local and do not depend on the
fermionic fields $\V_n$, $\bar{\V}_n$, $B_n$ and $\bar{B}_n$: They are
functions only of $\p_{n}$ and $\s_n$. We thus obtain three conditions, 
{ \setcounter{enumi}{\value{equation}} \addtocounter{enumi}{1}
\setcounter{equation}{0}
\renewcommand{\theequation}{\thesection.\theenumi\alph{equation}}
\begin{eqnarray}
 \bar{\V}_n\left[
 \left\{\g_5 , \tilde{D}\right\} - \frac{2}{\a}\tilde{D}\g_5 \tilde{D} 
 - i\ \d \tilde{D}
 \right]_{nm}\V_m 
 & = & 0 ,
 \label{2.14}    
\end{eqnarray}
\begin{eqnarray}
\d A_{X}^{(0)}= 0,
\label{2.15}
\end{eqnarray}
\begin{eqnarray}
 2i \sum_n{\rm Tr}\left(\g_5 - \g_5\frac{\tilde{D}}{\a}\right)_{nn} 
- \d J_{X} + \d  A_{X}^{\rm count} 
 = 0.
\label{2.16}
\end{eqnarray}
\setcounter{equation}{\value{enumi}}
}
\hspace{-13pt}
Here  $A_{X}$ is divided into two parts, 
$A_{X} = A_{X}^{(0)} + A_{X}^{\rm count}$: $A_{X}^{(0)}$ denotes a
chiral invariant action, and $A_{X}^{\rm count}$ a counter action.

The first eq.(\ref{2.14}) is a generalization of the GW relation for
free theory. It depends on the auxiliary fields and tells us how to
construct an exact chiral symmetry in the macroscopic theory.  The
second eq.(\ref{2.15}) can be used to fix an invariant potential of
the auxiliary fields.  The third eq.(\ref{2.16}) represents the anomaly
matching relation: Because of the absence of anomaly in the
microscopic theory, the macroscopic theory should be anomaly free. The
first term in the l.h.s of (\ref{2.16}) can be interpreted as the change
of the macroscopic path integral measure ${\cal D}\V{\cal D}\bar{\V}$
under the chiral transformations constructed below.  The second term $\d
J_{X}$ is the change of the macroscopic path integral measure of the
auxiliary fields ${\cal D}\p{\cal D}\s$ under (\ref{2.11}). If these two
contributions do not vanish, they must be eliminated by the
contributions from the counter term $\d A_{X}^{\rm count}$.

\section{Solutions to the GW relation and other WT identities}
Let us consider first the GW relation (\ref{2.14}) which reduces to   
\begin{eqnarray}
\left[\left\{\g_5 , \tilde{D}\right\} - \frac{2}{\a}\tilde{D}\g_5 \tilde{D} 
 - i\ \d \tilde{D}
 \right]_{nm}=0 ,
\label{4.1}
\end{eqnarray}
where $\tilde{D}_{nm}=D_{nm}+\d_{nm}X_{n}$ with $X_{n}=(i\g_{5}\p +
\s)_{n}$. We use here again the locality assumption that $\d X_{n}$
should be only a function of $X_{n}$.
Then, the GW relation (\ref{4.1}) can be decomposed
into contact terms and non-contact terms: 
\begin{eqnarray}
&{}& \left\{\g_5 ,X_n\right\} 
        - \frac{2}{\a} X_n\g_5 X_n
        - i\ \d X_n = 0 ,
 \label{4.2}\\ 
&{}&  \left[\left\{\g_5 ,D\right\} 
        - \frac{2}{\a} D\g_5 D - \frac{2}{\a} D\g_5 X
        - \frac{2}{\a} X\g_5 D
        - i\ \d D \right]_{nm}= 0 .
 \label{4.3} 
\end{eqnarray}
Since the $X$ commutes with $\g_{5}$\footnote{The matrix $\g_{5}$
satisfies $\g_{5}^{2}=1$.}, it immediately follows from
(\ref{4.2}) that
\begin{eqnarray}
 \left\{
 \begin{array}{ccl} 
  \d X_{n}&=& -2i\g_{5}X_{n}\left(1- \fa X_{n}\right)\\ 
  \d\p_n & = & 
    -2\s_n + \frac{2}{\a}\left(\s_n^2 - \p_n^2\right) \\
  \d\s_n & = & 
    2\p_n - \frac{4}{\a}\s_n\p_n 
 \end{array}
 \right. .
 \label{4.4}
\end{eqnarray}

In order to solve (\ref{4.3}), we make an ansatz for the Dirac operator:
\begin{eqnarray}
 \tilde{D} 
 & \equiv &
 D_{0} + \left(1+{\cal L}(D_{0})\right) X \left(1+{\cal R}(D_{0})\right) 
   \nonumber \\
 &  = & D + X ,\nonumber  \\
 D & = & D_{0} + {\cal L}(D_{0})\,X +X\,{\cal R}(D_{0})
         +{\cal L}(D_{0})\,X\,{\cal R}(D_{0}),  
\label{4.5}
\end{eqnarray}
where the $ D_{0}$ is the Dirac operator in the free theory. It
satisfies the original GW relation,
\begin{eqnarray}
 \{\g_5,D_{0}\} = \frac{2}{\a}D_{0}\g_{5}D_{0}.
\label{4.6}
\end{eqnarray}
Let us suppose that a solution for $D_0$ such as the Neuberger's type
\cite{neu1} is given. The functions 
${\cal L} ( D_{0})$ and ${\cal R}( D_{0})$ can be
fixed by substituting (\ref{4.5}) into (\ref{4.3}) and using (\ref{4.4})
for $\d X$. After a little algebra, we find that the Dirac operator,
\begin{eqnarray}
 \tilde{D} 
  &=& 
 D_{0} + \left(1-\fa{}D_0\right)~ X~
         \left(1-\fa{}D_0 \ 
            \frac{1-\fat \g_5{}D_0{}\g_5}{1-\fa \g_5{}D_0{}\g_5}\right), 
\label{4.10}
\end{eqnarray}
solves the GW relation (\ref{4.1}). 

Let us determine the potential for the auxiliary fields $A_{X}[\p,\s]$. 
We first consider (\ref{2.14}) to find a potential term
$A_{X}^{(0)}[\p,\s]$ which is invariant under the nonlinear chiral
transformations (\ref{4.4}) for $\pi$ and $\s$.  It turns out to be
\begin{eqnarray}
 A_{X}^{(0)}[\p,\s] & = &
  \sum_{n}h\left(\frac{\p_{n}^2 + \s_{n}^2}
     {1-\frac{2\s_n}{\a}+\frac{\p_{n}^2 + \s_{n}^2}{\a^2}}\right) , 
  \label{4.11}
\end{eqnarray}
where $h(x)$ is a function of $x$. 

We next consider (\ref{2.16}) to fix the counter action 
$A_{X}^{\rm count} = A_{X}^{(1)} + A_{X}^{(2)}$. These obey
\begin{eqnarray}
 \d A^{(1)}_{X}[\p,\s] &=& \d J_{X}[\p,\s] = -\sum_{n}\left(\frac{8}{\a}\p_n\right) \nonumber \\
     &=& i\frac{2^{2-{\frac{d}{2}}}}{\a}\sum_{n}{\rm Tr}\g_5(X-X^{\dag}),
  \label{4.12} \\
 \d A^{(2)}_{X}[\p,\s] &=& - 2i \sum_n{\rm Tr}\left(\g_5 - \g_5\frac{\tilde{D}}{\a}\right)_{nn}\nonumber\\
&=&
    i\fat \sum_n{\rm Tr}\left(\g_5(1+{\cal L})X(1+{\cal R})\right)_{nn} ,
  \label{4.13}
\end{eqnarray} 
where the trace ${\rm Tr}$ is taken over the spinor space.
Solving these, we find 
\begin{eqnarray}
 A_{X}^{(1)}[\p,\s] &=& \sum_{n}2
  \ln\left(1-\frac{2\s_n}{\a}+\frac{\p_{n}^2 + \s_{n}^2}{\a^2}\right)
\nonumber\\
  &=&   2^{1-\frac{d}{2}}\sum_{n}{\rm Tr}
   \ln\left(\left(1-\frac{X_n}{\a}\right)
      \left(1-\frac{X_n^{\dag}}{\a}\right)\right) , 
\label{4.14}  \\ \nn \\
 A_{X}^{(2)}[\p,\s] & = &
  \sum_{n}{\rm Tr}\left(\g_5(1+{\cal L})\g_5\ln\left(1-\frac{X}{\a}\right)
   \cdot
    (1+{\cal R})\right)_{nn}.   
\label{4.15}
\end{eqnarray}

We summarize here the structure of the chiral symmetry in the macroscopic
theory. We may define the chiral transformation\footnote{For the Dirac
fields, the most general form of the chiral transformation may be given by 
$\d \V =i\g_5 [1- \a^{-1}(2- s)\tilde{D}] \V,~~~\d \bar{\V} =  \bar{\V}i 
(1- \a^{-1}s \tilde{D}) \g_{5}$, where $s$ is a real constant. In
(\ref{5.1}), we take $s=0$.} as
\begin{eqnarray}
\d \V_{n} &=& i\g_5\left(1-\fat \tilde{D}\right)_{nm}\V_{m} \nonumber\\
\d \bar{\V}_{n}&=& \bar{\V}_{n}\ i\g_5 \nonumber\\
 \d X_{n} &=& -2i\g_5\left(X_{n}-\fa X_{n}X_{n}\right).
\label{5.1}
\end{eqnarray}
In the path-integral over the macroscopic fields, 
\begin{eqnarray}
 \lefteqn{
   Z_{\rm MACRO}} \nn \\
 & = & 
 \int 
\left[{\cal D}\V{\cal D}\bar{\V}{\cal D}\p{\cal D}\s\  
              e^{-A_{X}^{(1)}[\p,\s] -A_{X}^{(2)}[\p,\s]}
             \right]
 e^{-
\sum_{nm}\bar{\V}_n {\tilde D}(\p,\s)_{nm} \V_m
             -A_{X}^{(0)}[\p,\s]} ,
\label{5.2}
\end{eqnarray}
the measure ${\cal D}\V{\cal D}\bar{\V}{\cal D}\p{\cal D}\s$ multiplied
by the exponential of the counter actions $e^{-(A_{X}^{(1)} + A_{\rm
aux}^{(2)})}$ becomes invariant under (\ref{5.1}). The remaining action
\begin{eqnarray}
 A^{\rm inv}= \sum_{nm}\bar{\V}_n\left(D_0 
             + (1+{\cal L})(i\g_5\p +\s)(1+{\cal R})\right)_{nm}\V_m
             +A_{X}^{(0)}[\p,\s]
\label{5.3}
\end{eqnarray}
is also chiral invariant: $\d A^{\rm inv}=0$.  In summary, we construct an
exact chiral symmetry in quantum system (\ref{5.2}) with the auxiliary fields
describing the scalar and pseudo-scalar couplings of the fermionic
self-interactions.

\section{Summary and discussion}
In this letter, we have constructed an exact chiral symmetry in the
quantum system (\ref{5.2}) on the coarse lattice with the auxiliary
fields describing the scalar and pseudo-scalar couplings of the
fermionic self-interactions. The GW relation plays a key r\^{o}le in
this construction, as in the case of chiral gauge theories.  It should
be remarked, however, that the GW relation obtained here and the one for
chiral gauge theories have different structure: Although the Dirac
operator in the latter depends on the gauge field, its expression takes
the same form as that for free theories. In contrast to this, our
GW relation (\ref{4.1}) has additional contribution from the auxiliary
fields. We have solved the relation under the locality assumption on the
auxiliary fields. The solutions are obtained in a closed form, and used
to construct the chiral transformations of the Dirac as well as the
auxiliary fields on the coarse lattice. We have also solved other WT
identities to fix the auxiliary-field sector of the macroscopic action,
i.e., the invariant action and the counter action needed to define
chiral invariant measure. In this way we have obtained chiral symmetric
partition function (\ref{5.2}).

The lattice chiral symmetry discussed here has yet some peculiar
properties.  All the fields on the coarse lattice transform nonlinearly. 
The $\d \V$ and $\d {\bar \V}$ depend on $X$, and $\d X$ has $X^{2}$
term. As a result, none of the kinetic term $\bar{\V}D_0\V$, the Yukawa
coupling term $\bar{\V}(1+{\cal L})(i\g_{5}\p + \s)(1+{\cal R})\V$, the
functional measure ${\cal D}\V{\cal D}\bar{\V}{\cal D}\p{\cal D}\s$
becomes chiral invariant.  Furthermore, the invariant operators made out
of the auxiliary fields are non-polynomial, as seen in (\ref{4.11}). It
makes the integration of the auxiliary fields difficult. These problems
are related to the choice of the dynamical variables in our system. In a
forthcoming paper \cite{isu}, we describe our construction of the
symmetry in more detail, and discuss the above problems as well as the
structure of our lattice chiral symmetry.

\section*{Acknowledgments} 

This work is supported in part by the Grants-in-Aid for Scientific
Research No. 12640258, 12640259, and 13135209 from the Japan Society for
the Promotion of Science.



\vspace{0.5cm}
\begin{thebibliography}{99}
\bibliographystyle{unsrt} 
\setlength{\itemsep}{0.0in}
 

\bibitem{gw} P.~H.~Ginsparg and K.~G.~Wilson, Phys.\ Rev.\ {\bf D25}
	(1982) 2649. 
\bibitem{neu1} H.~Neuberger, Phys.\ Lett.\ {\bf B417} (1998) 141; Phys.\
		Lett.\ {\bf B427} (1998) 353.
\bibitem{hln}  P.~Hasenfratz, V.~Laniena and F.~Nidermayer, Phys.\ Lett.\ {\bf B427} (1998) 125.
\bibitem{lu1} M.~L{\"u}scher, Phys.\ Lett.\ {\bf B428} (1998) 342.
\bibitem{has} P.~Hasenfratz, Nucl.\ Phys.\ {\bf B525} (1998) 401.  
\bibitem{her} P.~Hernandez, K.~Jansen and M.~L{\"u}scher, 
Nucl.\ Phys.\ {\bf B552} (1999) 363.  
\bibitem{neu2}  H.~Neuberger, Phys.\ Rev.\ {\bf D57} (1998) 5417.
\bibitem{Hor} I.~Horvath, Phys. Rev. Lett.\  {\bf 81} (1998) 4063. 
\bibitem{nn}  H.~B.~Nielsen and M.~Ninomiya,
 Nucl.\ Phys.\ {\bf B185} (1981) 20;
 [E: {\bf B195} (1982) 541]; {\bf B193} (1981) 173.
\bibitem{reviews} F.~Nidermayer, Nucl.\ Phys.\ Proc. \ Supple. \ {\bf
		73} (1999) 105.\\
H.~Neuberger, ``Exact chiral symmetry on the lattice'', hep-lat/0101006.\\
M.~L{\"u}scher, ``Chiral gauge theories revised'', hep-th/0102028.\\
M.~Creutz, ``Aspects of chiral symmetry and the lattice'', hep-lat/0007032.
\bibitem{lu2} M.~L{\"u}scher,  Nucl.\ Phys.\ {\bf B538} (1999) 515;  Nucl.\ Phys.\ {\bf B549} (1999) 295. 
\bibitem{cwz} T.~W.~Chiu, C.~W.~Wang and S.~V.~Zenkin,
             Phys.\ Lett.\ {\bf B438} (1998) 321. 
\bibitem{in} I.~Ichinose and K.~Nagao, Chin.\ J.\ Phys.\ {\bf 38} (2000)
	671; hep-lat/9909035; Mod.\ Phys.\ Lett.\ {\bf A15} (2000) 857. 
\bibitem{iis} Y.~Igarashi, K.~Itoh and H.~So, Phys.\ Lett.\ {\bf B526} 
(2002) 164. 
\bibitem{isu}  Y.~Igarashi, H.~So and N.~Ukita, in preparation.


\end{thebibliography}
\end{document}